

\newif\ifproofmode			
\proofmodefalse				

\newif\ifforwardreference		
\forwardreferencetrue			

\newif\ifeqchapternumbers		
\eqchapternumbersfalse			

\newif\ifsectionnumbers			
\sectionnumberstrue			

\newif\ifeqsectionnumbers		
\eqsectionnumbersfalse			

\newif\ifchaptersectionnumbers     	
\chaptersectionnumberstrue		

\newif\ifcontinuoussectionnumbers	
\continuoussectionnumbersfalse	

\newif\ifcontinuousnumbers		
\continuousnumbersfalse 		

\newif\iffigurechapternumbers		
\figurechapternumbersfalse		

\newif\ifcontinuousfigurenumbers	
\continuousfigurenumbersfalse		

\newif\ifcontinuousreferencenumbers     
\continuousreferencenumberstrue         

\newif\ifparenequations			
\parenequationstrue			

\newif\ifcrossreference			
\crossreferencefalse			

\newif\ifstillreading			

\font\eqsixrm=cmr6			
\def\marginstyle{\eqsixrm}		

\newtoks\chapletter			
\newcount\chapno			
\newcount\sectno			
\newcount\eqlabelno			
\newcount\figureno			
\newcount\referenceno			
\newcount\minutes			
\newcount\hours				

\newread\labelfile			
\newwrite\labelfileout			
\newwrite\allcrossfile			

\chapno=0
\sectno=0
\eqlabelno=0
\figureno=0


\def\chapternumberstrue{\eqchapternumberstrue}

%
\def\initialeqmacro{
    \ifproofmode
        \headline{\tenrm \today\ --\ \timeofday\hfill
                         \jobname\ --- draft\hfill\folio}
        \hoffset=-1cm
        \ifcrossreference
            \immediate\openout\allcrossfile=zallcrossreferfile
        \fi
    \else
        \crossreferencefalse
    \fi
    \ifforwardreference
        \openin\labelfile=zlabelfile
        \ifeof\labelfile
        \else
            \stillreadingtrue
            \loop
                \read\labelfile to \nextline
                \ifeof\labelfile
                    \stillreadingfalse
                \else
                    \nextline
                \fi
                \ifstillreading
            \repeat
        \fi
        \immediate\openout\labelfileout=zlabelfile
    \fi}


{\catcode`\^^I=9
\catcode`\ =9
\catcode`\^^M=9
\endlinechar=-1
\globaldefs=1


%
\def\chapfolio{			
    \ifnum \chapno>0 \relax
        \the\chapno
    \else
        \the\chapletter
    \fi}

%
\def\bumpchapno{
    \ifnum \chapno>-1 \relax
        \global \advance \chapno by 1
    \else
        \global \advance \chapno by -1 \setletter\chapno
    \fi
    \ifcontinuousnumbers
    \else
        \global\eqlabelno=0
    \fi
    \ifcontinuousfigurenumbers
    \else
        \global\figureno=0
    \fi
    \ifcontinuousreferencenumbers
    \else
        \global\referenceno=0
    \fi
    \sectno=0}

\def\bumpsectno{
    \global\advance\sectno by 1 \relax
    \ifeqsectionnumbers
        \ifcontinuoussectionnumbers
        \else
            \global\eqlabelno=0
        \fi
    \fi}

%
\def\setletter#1{\ifcase-#1 {}  \or\global\chapletter={A}
  \or\global\chapletter={B} \or\global\chapletter={C} \or\global\chapletter={D}
  \or\global\chapletter={E} \or\global\chapletter={F} \or\global\chapletter={G}
  \or\global\chapletter={H} \or\global\chapletter={I} \or\global\chapletter={J}
  \or\global\chapletter={K} \or\global\chapletter={L} \or\global\chapletter={M}
  \or\global\chapletter={N} \or\global\chapletter={O} \or\global\chapletter={P}
  \or\global\chapletter={Q} \or\global\chapletter={R} \or\global\chapletter={S}
  \or\global\chapletter={T} \or\global\chapletter={U} \or\global\chapletter={V}
  \or\global\chapletter={W} \or\global\chapletter={X} \or\global\chapletter={Y}
  \or\global\chapletter={Z}\fi}

%
\def\tempsetletter#1{\ifcase-#1 {}\or{} \or\chapletter={A} \or\chapletter={B}
 \or\chapletter={C} \or\chapletter={D} \or\chapletter={E}
  \or\chapletter={F} \or\chapletter={G} \or\chapletter={H}
   \or\chapletter={I} \or\chapletter={J} \or\chapletter={K}
    \or\chapletter={L} \or\chapletter={M} \or\chapletter={N}
     \or\chapletter={O} \or\chapletter={P} \or\chapletter={Q}
      \or\chapletter={R} \or\chapletter={S} \or\chapletter={T}
       \or\chapletter={U} \or\chapletter={V} \or\chapletter={W}
        \or\chapletter={X} \or\chapletter={Y} \or\chapletter={Z}\fi}

%
\def\chapshow#1{
    \ifnum #1>0 \relax
        #1
    \else
        {\tempsetletter{\number#1}\the\chapletter}
    \fi}

%
\def\today{\number\day\space \ifcase\month\or Jan\or Feb\or
        Mar\or Apr\or May\or Jun\or Jul\or Aug\or Sep\or
        Oct\or Nov\or Dec\fi, \space\number\year}

\def\timeofday{\minutes=\time    \hours=\time
        \divide \hours by 60
        \multiply \hours by 60
        \advance \minutes by -\hours
        \divide \hours by 60
        \ifnum\the\minutes>9 \relax
     		\the\hours:\the\minutes
 	\else
  		\the\hours:0\the\minutes
	\fi}


%
%
%
%
\def\chapnum{\bumpchapno \chapfolio}

\def\chaplabel#1{
    \ifforwardreference                             
        \write\labelfileout{                        
        \noexpand\expandafter\noexpand\def          
        \noexpand\csname CHAPLABEL#1\endcsname{\the\chapno}}
    \fi
    \global\expandafter\edef\csname CHAPLABEL#1\endcsname
    {\the\chapno}
    \ifproofmode
        \rlap{\hbox{\marginstyle #1\ }}
    \fi}

%
\def\sectnum{
    \bumpsectno
        \ifchaptersectionnumbers
            \chapfolio.
        \fi
    \the\sectno}

\def\sectlabel#1{
    \bumpsectno
    \ifforwardreference
        \immediate\write\labelfileout{
        \noexpand\expandafter\noexpand\def
        \noexpand\csname SECTLABEL#1\endcsname{\the\chapno.\the\sectno?!}}
    \fi
    \global\expandafter\edef\csname SECTLABEL#1\endcsname
    {\the\chapno.\the\sectno?!}	 			
    \ifproofmode
        \llap{\hbox{\marginstyle #1\ }}
    \fi
    \ifchaptersectionnumbers
        \chapfolio.
    \fi
    \the\sectno}

\def\sectref#1{                                  
    \ifundefined{SECTLABEL#1}                     
        ++                                        
        \ifproofmode
            \ifforwardreference
            \else
            \write16{ ***Undefined\space Section\space Reference\space #1*** }
            \fi
        \else
        \write16{ ***Undefined\space Section\space Reference\space #1*** }
        \fi
    \else
        \edef\LABxx{\getlabel{SECTLABEL#1}}
	\ifchaptersectionnumbers
            \def\LAByy{\expandafter\stripchap\LABxx}
	    \chapshow\LAByy.
	\fi
	\expandafter\stripsect\LABxx
    \fi
    \ifcrossreference
        \write\allcrossfile{Section\space #1}
    \fi}

%
%
\def\eqnum{                                    
    \global\advance\eqlabelno by 1              
    \eqno(
    \ifeqchapternumbers
        \chapfolio.
    \fi
    \ifeqsectionnumbers
        \the\sectno.
    \fi
    \the\eqlabelno)}

\def\eqlabel#1{                                
    \global\advance\eqlabelno by 1              
    \ifforwardreference                     
        \immediate\write\labelfileout{\noexpand\expandafter\noexpand\def
        \noexpand\csname EQLABEL#1\endcsname
        {\the\chapno.\the\sectno?\the\eqlabelno!}}
    \fi
    \global\expandafter\edef\csname EQLABEL#1\endcsname
    {\the\chapno.\the\sectno?\the\eqlabelno!}
    \eqno(
    \ifeqchapternumbers
        \chapfolio.
    \fi
    \ifeqsectionnumbers
        \the\sectno.
    \fi
    \the\eqlabelno)
    \ifproofmode
        \rlap{\hbox{\marginstyle #1}}		
    \fi}

\def\eqalignnum{                               
    \global\advance\eqlabelno by 1              
    &(\ifeqchapternumbers
        \chapfolio.
    \fi
    \ifeqsectionnumbers
        \the\sectno.
    \fi
    \the\eqlabelno)}

\def\eqalignlabel#1{                   	
    \global\advance\eqlabelno by 1 	        
    \ifforwardreference                     
        \immediate\write\labelfileout{\noexpand\expandafter\noexpand\def
        \noexpand\csname EQLABEL#1\endcsname
        {\the\chapno.\the\sectno?\the\eqlabelno!}}
    \fi
    \global\expandafter\edef\csname EQLABEL#1\endcsname
    {\the\chapno.\the\sectno?\the\eqlabelno!}
    &(\ifeqchapternumbers
        \chapfolio.
    \fi
    \ifeqsectionnumbers
        \the\sectno.
    \fi
    \the\eqlabelno)
    \ifproofmode
        \rlap{\hbox{\marginstyle #1}}			
    \fi}

\def\dnum{                                     
    \global\advance\eqlabelno by 1              
    \llap{(	 				
    \ifeqchapternumbers
        \chapfolio.
    \fi
    \ifeqsectionnumbers
        \the\sectno.
    \fi
    \the\eqlabelno)}}

\def\dlabel#1{                                 
    \global\advance\eqlabelno by 1              
    \ifforwardreference                         
        \immediate\write\labelfileout{\noexpand\expandafter\noexpand\def
        \noexpand\csname EQLABEL#1\endcsname
        {\the\chapno.\the\sectno?\the\eqlabelno!}}
    \fi
    \global\expandafter\edef\csname EQLABEL#1\endcsname
    {\the\chapno.\the\sectno?\the\eqlabelno!}
    \llap{(
    \ifeqchapternumbers
        \chapfolio.
    \fi
    \ifeqsectionnumbers
        \the\sectno.
    \fi
    \the\eqlabelno)}
    \ifproofmode
        \rlap{\hbox{\marginstyle #1}}		
    \fi}

\def\eqref#1{\ifparenequations(\fi
    \ifundefined{EQLABEL#1}***
        \ifproofmode
            \ifforwardreference
            \else
            \write16{
                ***Undefined\space Equation\space Reference\space #1*** }
            \fi
        \else
        \write16{ ***Undefined\space Equation\space Reference\space #1*** }
        \fi
    \else
        \edef\LABxx{\getlabel{EQLABEL#1}}
	\def\LAByy{\expandafter\stripsect\LABxx}
        \def\LABzz{\expandafter\stripchap\LABxx}
        \ifeqchapternumbers
            \chapshow{\LABzz}.
        \else
            \ifnum \number\LABzz=\chapno \relax
            \else
                \chapshow{\LABzz}.
            \fi
        \fi
	\ifeqsectionnumbers
	    \LAByy.
	\fi
        \expandafter\stripeq\LABxx
    \fi
    \ifparenequations)\fi
    \ifcrossreference
        \write\allcrossfile{Equation\space #1}
    \fi}

%
\def\fignum{                                   
    \global\advance\figureno by 1\relax         
    \iffigurechapternumbers
        \chapfolio.
    \fi
    \the\figureno}

\def\figlabel#1{				
    \global\advance\figureno by 1\relax 	
    \ifforwardreference				
        \immediate\write\labelfileout{\noexpand\expandafter\noexpand\def
        \noexpand\csname FIGLABEL#1\endcsname
        {\the\chapno.\the\sectno?\the\figureno!}}
    \fi
    \global\expandafter\edef\csname FIGLABEL#1\endcsname
    {\the\chapno.\the\sectno?\the\figureno!}
    \iffigurechapternumbers
        \chapfolio.
    \fi
    \ifproofmode
        \llap{\hbox{\marginstyle #1\ }}\relax
    \fi
    \the\figureno}

\def\figref#1{					
    \ifundefined				
        {FIGLABEL#1}!!!!			
        \ifproofmode
            \ifforwardreference
            \else
            \write16{
                ***Undefined\space Figure\space Reference\space #1*** }
            \fi
        \else
        \write16{ ***Undefined\space Figure\space Reference\space #1*** }
        \fi
    \else
        \edef\LABxx{\getlabel{FIGLABEL#1}}
        \def\LABzz{\expandafter\stripchap\LABxx}
        \iffigurechapternumbers
            \chapshow{\LABzz}.\expandafter\stripeq\LABxx
        \else \ifnum\number\LABzz=\chapno \relax
                \expandafter\stripeq\LABxx
            \else
                \chapshow{\LABzz}.\expandafter\stripeq\LABxx
            \fi
        \fi
        \ifcrossreference
            \write\allcrossfile{Figure\space #1}
        \fi
    \fi}

%
%
\def\pagelabel#1{
    \ifforwardreference
        \write\labelfileout{
        \noexpand\expandafter\noexpand\def
        \noexpand\csname PGLABEL#1\noexpand\endcsname{\the\pageno}}
    \fi
    \global\expandafter\edef\csname PGLABEL#1\endcsname{\the\pageno}}

\def\pageref#1{
    \ifundefined
        {PGLABEL#1}***
        \ifproofmode
        \else
        \write16{ ***Undefined\space Page\space Reference\space #1*** }
        \fi
    \else
        \csname PGLABEL#1\endcsname
    \fi
    \ifcrossreference
        \write\allcrossfile{Page\space #1}
    \fi}

%
\def\refnum{                                      
    \global\advance\referenceno by 1\relax         
    \the\referenceno}	                           

\def\internalreflabel#1{			
    \global\advance\referenceno by 1\relax 	
    \ifforwardreference				
        \immediate\write\labelfileout{\noexpand\expandafter\noexpand\def
        \noexpand\csname REFLABEL#1\endcsname
        {\the\chapno.\the\sectno?\the\referenceno!}}
    \fi
    \global\expandafter\edef\csname REFLABEL#1\endcsname
    {\the\chapno.\the\sectno?\the\figureno!}
    \ifproofmode
        \llap{\hbox{\marginstyle #1\hskip.5cm}}\relax
    \fi
    \the\referenceno}

\def\internalrefref#1{				
    \ifundefined				
        {REFLABEL#1}!!!!			
        \ifproofmode
            \ifforwardreference
            \else
            \write16{
                 ***Undefined\space Footnote\space Reference\space #1*** }
            \fi
        \else
        \write16{
             ***Undefined\space Footnote\space Reference\space #1*** }
        \fi
    \else
        \edef\LABxx{\getlabel{REFLABEL#1}}
        \def\LABzz{\expandafter\stripchap\LABxx}
        \expandafter\stripeq\LABxx
        \ifcrossreference
            \write\allcrossfile{Reference\space #1}
        \fi
    \fi}

%
\def\reflabel#1{\item{\internalreflabel{#1}.}}

%
\def\refref#1{\internalrefref{#1}}

\def\eq{\ifhmode Eq.~\else Equation~\fi}		
\def\eqs{\ifhmode Eqs.~\else Equations~\fi}

%
%
%
%

%
\def\getlabel#1{\csname#1\endcsname}
\def\ifundefined#1{\expandafter\ifx\csname#1\endcsname\relax}
\def\stripchap#1.#2?#3!{#1}			
\def\stripsect#1.#2?#3!{#2}			%
\def\stripeq#1.#2?#3!{#3}			
}  

\overfullrule = 0pt
\magnification = 1200
\baselineskip 21pt plus 0.2 pt minus 0.2 pt
\chapternumberstrue
\initialeqmacro
\def\sh{\mathop{\rm sh}\nolimits}
\def\ch{\mathop{\rm ch}\nolimits}

\line{\hfill UMTG-177}

\vskip 0.2 in

\centerline{\bf Direct Calculation of the Boundary $S$ Matrix}

\centerline{\bf for the Open Heisenberg Chain}

\bigskip

\medskip

\centerline{M. T. Grisaru\footnote{$\dagger$}{Physics Department,
Brandeis University, Waltham, MA 02254},
Luca Mezincescu${}^*$, and
Rafael I. Nepomechie\footnote*{Department of Physics, University of Miami,
Coral Gables, FL 33124}}

\vskip 0.2 in

\bigskip

\centerline{\bf Abstract}

We calculate the boundary $S$ matrix for the open antiferromagnetic
spin $1/2$ isotropic Heisenberg chain with boundary magnetic fields.
Our approach, which starts from the model's Bethe Ansatz solution,
is an extension of the Korepin-Andrei-Destri method. Our result
agrees with the boundary $S$ matrix for the boundary sine-Gordon model
with $\beta^2 \rightarrow 8\pi$ and with ``fixed'' boundary conditions.

\vskip 0.2 in

\vfill\eject

\noindent
{\bf \chapnum . Introduction}
\vskip 0.2truein

In a $1+1$-dimensional theory with factorized scattering,
the two-particle $S$ matrix, which depends on the rapidity difference
$\lambda$ of the two particles and which we denote here by $R(\lambda)$,
is constrained to satisfy the factorization (Yang-Baxter) equation
$$
R_{12}(\lambda - \lambda')\  R_{13}(\lambda)\ R_{23}(\lambda') =
R_{23}(\lambda')\  R_{13}(\lambda)\ R_{12}(\lambda - \lambda')
\,.  \eqlabel{yang-baxter}
$$
For particles in an $n$-dimensional representation of some internal symmetry
group, $R(\lambda)$ is a matrix acting in the tensor product space
$C^n \otimes C^n$. Moreover,
$R_{12}$, $R_{13}$, and $R_{23}$ are matrices
acting in $C^n \otimes C^n \otimes C^n$, with
$R_{12} = R \otimes 1$,
$R_{23} = 1 \otimes R$, etc. (See, e.g., Refs.
\refref{zamolodchikov/zamolodchikov}, \refref{reviews} and references therein.)

For a system with a boundary,
the quantum-mechanical scattering of a particle with the boundary is described
by a so-called boundary $S$ matrix, which we denote here by
$K(\lambda \,, \xi)$. This is a matrix acting in the space $C^n$, which may
depend on one or more parameters, which we denote collectively by $\xi$,
that characterize the interaction at the boundary.
The condition that boundary scattering be
compatible with factorization is${}^{\refref{cherednik} -
\refref{nonsymmetric}}$
$$ R_{12}(\lambda - \lambda')\ K_1(\lambda \,, \xi)\
R_{21}(\lambda + \lambda')\ K_2(\lambda' \,, \xi)
= K_2(\lambda'\,, \xi)\ R_{12}(\lambda + \lambda')\ K_1(\lambda\,, \xi)\
R_{21}(\lambda - \lambda') \,,
\eqlabel{reflection} $$
where
$$R_{21}(\lambda) \equiv {\cal P}_{12}\ R_{12}(\lambda)\ {\cal P}_{12}
\,,  \eqnum $$
and ${\cal P}_{12}$ is the permutation matrix in $C^n \otimes C^n$,
$${\cal P}_{12}  (x \otimes y) = y \otimes x \quad {\hbox{for}} \quad
x,y \in C^n \,. \eqlabel{permutation} $$
Furthermore, we use the notation $K_1 = K \otimes 1$, $K_2 = 1 \otimes K$.

Given $R(\lambda)$, \eq\eqref{reflection} determines the general form of
$K(\lambda \,, \xi)$ up to a scalar factor, which is a function of $\lambda$
and $\xi$.  Such
boundary $S$ matrices have been used to construct and to obtain the
Bethe-Ansatz solution for {\it open} integrable quantum spin
chains${}^{\refref{sklyanin} - \refref{ruiz}}$.
For these applications, what is important is the matrix structure of
$K(\lambda \,, \xi)$; the scalar factor plays no significant role.
Further applications of boundary $S$ matrices are reviewed in
Ref. \refref{kulish}.

Recently, there has been
interest${}^{\refref{fring/koberle} - \refref{chim}}$ in
determining
the boundary $S$ matrices for physical excitations of integrable
theories. This requires determining in particular the scalar factor of the
boundary $S$ matrix, which contains information about possible boundary bound
states, etc.

One way to determine the scalar factor is the ``bootstrap'' approach:
in addition to \eq\eqref{reflection}, one imposes on the boundary $S$ matrix
the constraints of unitarity,
boundary cross-unitarity${}^{\refref{ghoshal/zamolodchikov}}$, and
boundary bootstrap${}^{\refref{fring/koberle},
\refref{ghoshal/zamolodchikov}}$ (which in the context of spin chains
is known as ``fusion'' of boundary $S$ matrices${}^{\refref{spin1/fusion}}$).
These conditions determine the scalar factor up to a CDD-type of ambiguity.
Evidently, this procedure is a generalization of the bootstrap
approach for determining bulk $S$
matrices${}^{\refref{zamolodchikov/zamolodchikov}}$.
Ghoshal and Zamolodchikov${}^{\refref{ghoshal/zamolodchikov}}$ have
determined in this manner the boundary $S$ matrix for the boundary
sine-Gordon model.

In this paper, we consider an alternative approach: namely, to determine
the boundary $S$ matrix directly from the ``microscopic'' theory
for the excitations -- i.e., from the Bethe Ansatz equations.
In particular, we calculate in this way the boundary
$S$ matrix for the physical excitations (``spinons'') of the open
antiferromagnetic spin $1/2$ Heisenberg chain with boundary magnetic fields.
The Hamiltonian of the chain is given by
$${\cal H} = {1\over 4}\left\{ \sum_{n=1}^{N-1} \vec \sigma_n \cdot
\vec \sigma_{n+1} + {1\over \xi_-}\sigma^z_1 + {1\over \xi_+}\sigma^z_N
\right\} \,, \eqlabel{hamiltonian}
$$
where $\vec\sigma$ are the usual Pauli matrices, and the real parameters
$\xi_\pm$ correspond to boundary magnetic fields.
Although the bulk terms are $SU(2)$ invariant, the boundary terms
break this symmetry down to $U(1)$, which is generated by
$$S^z = {1\over 2}\sum_{n=1}^N \sigma^z_n \,. \eqnum $$
The Bethe Ansatz solution of this model is given in Refs.
\refref{sklyanin} and \refref{alcaraz}.

Our approach is a generalization to the case of systems with boundaries
of the Korepin-Andrei-Destri method
${}^{\refref{korepin}, \refref{andrei/destri}}$, which was devised
to calculate factorized bulk $S$ matrices for systems with
periodic boundary conditions.
A key ingredient is the quantization condition for a finite interval,
which has recently been discussed by Fendley and
Saleur${}^{\refref{fendley/saleur}}$. Our result $K(\lambda, \xi_\pm)$
for the boundary $S$ matrix is the diagonal matrix
(see Eqs. \eqref{form},\eqref{result1}, \eqref{result2})
$$K(\lambda, \xi_\pm) = \alpha(\lambda, \xi_\pm)
\left( \matrix{1  & 0 \cr
               0  & -{\lambda +i(\xi_\pm -{1\over 2})\over
\lambda -i(\xi_\pm -{1\over 2})} \cr}\right) \,,
\eqlabel{resulta} $$
where the scalar factor $\alpha(\lambda, \xi_\pm)$ is given by
$$\alpha(\lambda, \xi_\pm) =
{\Gamma \left({-i\lambda\over 2} + {1\over 4}\right) \over
  \Gamma \left({i\lambda\over 2} + {1\over 4}\right)}
{\Gamma \left({i\lambda\over 2} + 1\right) \over
\Gamma \left({-i\lambda\over 2} + 1\right)}
{\Gamma \left({-i\lambda\over 2} + {1\over 4}(2\xi_\pm -1)\right)\over
  \Gamma \left({i\lambda\over 2} + {1\over 4}(2\xi_\pm -1)\right)}
{\Gamma \left({i\lambda\over 2} + {1\over 4}(2\xi_\pm +1)\right)\over
\Gamma \left({-i\lambda\over 2} + {1\over 4}(2\xi_\pm +1)\right)} \,.
\eqlabel{resultb} $$

Let us compare the above result with previous field theory results.
The bulk $S$ matrix for the Heisenberg chain${}^{\refref{faddeev/takhtajan}}$
coincides with the bulk $S$ matrix for the sine-Gordon
model${}^{\refref{zamolodchikov/zamolodchikov}}$ in the limit
$\beta^2 \rightarrow 8\pi$, which is the $SU(2)$-invariant point.
\footnote*{More precisely, the {\it true} $S$ matrices
$\check R = {\cal P} R$ for the two models are related by a unitary
transformation.}
Therefore, we expect that the boundary $S$ matrix for the open Heisenberg chain
with boundary magnetic fields \eqref{resulta}, \eqref{resultb}
should coincide with the boundary $S$ matrix
of Ghoshal and Zamolodchikov${}^{\refref{ghoshal/zamolodchikov}}$
for the boundary sine-Gordon model with $\beta^2 \rightarrow 8\pi$ and with
``fixed'' boundary conditions. (For ``fixed'' boundary conditions,
the field theory and hence the boundary $S$ matrix are $U(1)$ invariant.)
We have verified that the two boundary $S$ matrices indeed coincide,
up to a rapidity-independent scalar factor, and with some redefinitions
of variables.\footnote{$\dagger$}{In the notation of
Ref. \refref{ghoshal/zamolodchikov}, ``fixed'' boundary
conditions corresponds to the case $k=0$. Moreover, the isotropic limit
is performed by rescaling $\xi \rightarrow \lambda \xi - {\pi\over 2}$ (here
$\lambda \equiv (8\pi/\beta^2) - 1$), and taking the limit
$\lambda \rightarrow 0$ with rapidity $u = -i \theta$ fixed. Our result
for the boundary $S$ matrix is recovered by making the identifications
$\theta = \pi \lambda$ and $\xi = \pi ( \xi_\pm - {1\over 2})$.}

The bootstrap result of Ghoshal and Zamolodchikov for the boundary
sine-Gordon model with ``fixed'' boundary conditions has been verified
using the physical Bethe Ansatz approach by Fendley and
Saleur${}^{\refref{fendley/saleur}}$.

We now outline the contents of the paper. In Section 2, we consider
the problem of quantization on a finite interval.
We begin by demonstrating that even the simplest
of systems can exhibit interesting boundary $S$ matrices. Indeed,
we reanalyze an example which is familiar to every student of
quantum mechanics -- a free nonrelativistic particle -- but with
nonstandard boundary conditions. This leads to a boundary $S$ matrix
which is momentum dependent, and which has a pole which may correspond
to a boundary bound state. We work out the quantization condition
on a finite interval, and then discuss the generalization for
factorized scattering. As already noted, this quantization condition
is a key ingredient of the calculation of boundary $S$ matrices
from the Bethe Ansatz.
In Section 3 we analyze the Bethe Ansatz equations for the model
\eqref{hamiltonian} using the string hypothesis.
In Section 4, we study the ground state. In particular, we compute
the root density to order $1/N$, and calculate the surface energy.
Although these results have previously been obtained
${}^{\refref{hamer/quispel/batchelor}}$, the calculations serve
as a useful preparation for the study of excited states.
Section 5 is the core of the paper. There we compute the density
of roots and holes to order $1/N$ for excited states. With the
help of the quantization condition, we then determine the boundary $S$
matrix. We also perform a nontrivial consistency check on our result.
There is a brief discussion in Section 6.
At several points in Sections 4 and 5, we must approximate certain
sums by integrals, being careful to keep terms of order $1/N$.
We derive an appropriate formula in the Appendix with the help of
the Euler-Maclaurin formula.

\vfill\eject

\vskip 0.4truein
\noindent
{\bf \chapnum . Quantization on a finite interval}
\vskip 0.2truein

\noindent
{\it Nonrelativistic scattering}
\vskip 0.2truein

As a warm-up exercise, we first
consider a free 1-dimensional nonrelativistic particle of mass $m$
on the half-line $x \ge 0$. Usually one demands that the wavefunction
$\psi(x)$ vanish at $x=0$. This
is a sufficient, but by no means necessary, condition for the probability
current $j(x) = i \psi(x)^* {\buildrel \leftrightarrow \over \partial_x}
\psi(x)$ to vanish at $x=0$. We consider
instead the more general (mixed Dirichlet-Neumann) boundary condition
$$ c\psi(x) + {d\over dx}\psi(x) = 0 \quad\quad {\hbox{  at  }} x=0 \,,
\eqlabel{bc} $$
where $c$ is a real parameter of dimension 1/length. This boundary condition
also implies the
vanishing of the probability current at $x=0$, and is compatible with
the self-adjointness of the Hamiltonian $H=p^2/2m$.
Assuming energy eigenfunctions of the plane-wave form
$$\psi_p(x) = A e^{i p x} + B e^{-i p x}   \eqlabel{planewave} $$
(we set $\hbar = 1$), we can use the boundary condition \eqref{bc} to
eliminate $A$ in terms of $B$; and we immediately obtain
$$\psi_p(x) = B \left[ e^{-i p x} + \left( {p + ic\over p -ic} \right)
e^{i p x} \right] \,. \eqnum $$
We conclude that the boundary $S$ matrix is given by
$$K(p) = {p + ic\over p -ic} \,. \eqnum $$
The pole at $p=ic$ implies the existence (for $c > 0$ ) of a boundary
bound state with energy $E = -c^2/2m$.

We remark that for $c=0$, we have the Neumann boundary condition
$\psi'(0) = 0$, and the wavefunction is
$$\psi_p(x) \sim \cos px \,; \eqnum $$
while for $c \rightarrow \infty$, we have the Dirichlet boundary condition
$\psi(0) = 0$, and the wavefunction is
$$\psi_p(x) \sim \sin px \,. \eqnum $$
Wavefunctions with such properties have appeared in quantum impurity
problems. (See, e.g., Ref. \refref{affleck}.)

We next consider the problem of a free particle on the finite interval
$-L/2 \le x \le L/2$, with mixed Dirichlet-Neumann boundary conditions
at both ends:
$$ \mp c_\pm \psi(x) + {d\over dx}\psi(x) = 0 \quad\quad {\hbox{  at  }}
x=\pm {L\over 2} \,.
\eqlabel{bcinterval} $$
Assuming again that the energy eigenfunctions are plane waves
\eqref{planewave}, and imposing the above boundary conditions, we see that
$p$ obeys the quantization condition
$$e^{i 2 p L}\ K^-(p)\ K^+(p) = 1 \,, \eqlabel{q1} $$
where the boundary $S$ matrices $K^\pm$ are given by
$$K^\pm (p) = {p + ic_\pm\over p -ic_\pm} \,. \eqnum $$
Of course, the momentum operator is not defined on a finite interval,
and therefore $p$ is not a momentum eigenvalue. Nevertheless, the energy
is still given by $E=p^2/2m$.

Finally, if we now turn on a reflectionless potential which is localized near
$x=0$, the quantization condition \eqref{q1} is generalized to
$$e^{i 2 p L}\ R(p)^2\ K^-(p)\ K^+(p) = 1 \,, \eqlabel{q2} $$
where $R(p)$ is the $S$ matrix for the potential. We shall now see
that this formula has a straightforward generalization for factorized
scattering.

\vfill\eject

\noindent
{\it Factorized scattering}
\vskip 0.2truein

Consider a system of particles with factorized scattering
on an interval of finite length $L$.
Each particle has some rapidity $\lambda$, and in general also carries
internal quantum numbers, such as spin. We denote the energy of a particle
by $\varepsilon (\lambda)$, and we define $p(\lambda)$ by the
expression for the momentum of a particle for the corresponding system with
periodic boundary conditions. For the case of
two such particles, with corresponding rapidities $\lambda_1$ and $\lambda_2$,
the quantization condition reads (in the notation used in
Eqs.\eqref{yang-baxter} - \eqref{permutation})
$$e^{i 2 p(\lambda_1) L}\ R_{12}(\lambda_1 - \lambda_2)\
K_1(\lambda_1 \,, \xi_-)\
R_{21}(\lambda_1 + \lambda_2)\ K_1(\lambda_1 \,, \xi_+) = 1 \,. \eqlabel{q3} $$
This quantization condition has recently been discussed in
Ref. \refref{fendley/saleur}. As noted there, such a formula can be
derived with the help of the Zamolodchikov-Faddeev algebra.

\vskip 0.4truein
\noindent
{\bf \chapnum . Bethe Ansatz and string hypothesis}
\vskip 0.2truein

The Hamiltonian ${\cal H}$ for the open antiferromagnetic Heisenberg chain
with boundary magnetic fields is given by \eq\eqref{hamiltonian}. The
simultaneous eigenstates of ${\cal H}$ and $S^z$ have been determined by both
the coordinate${}^{\refref{alcaraz}}$ and algebraic${}^{\refref{sklyanin}}$
Bethe Ansatz. In the latter approach, one constructs certain creation
and destruction operators, ${\cal B}(\lambda)$ and ${\cal C}(\lambda)$,
respectively; and the eigenstates are given by
$${\cal B}(\lambda_1)\ {\cal B}(\lambda_2) \cdots {\cal B}(\lambda_M)\
\omega^+ \,, \eqlabel{state} $$
where $\omega^+$ is the ferromagnetic vacuum state with all spins up,
$$ {\cal C}(\lambda)\ \omega^+ = 0 \,, \eqlabel{pseudovacuum} $$
and $\{ \lambda_\alpha \}$ satisfy the Bethe Ansatz (BA) equations
$$\eqalignno{
\left({\lambda_\alpha + i(\xi_+ -{1\over 2}) \over
       \lambda_\alpha - i(\xi_+ -{1\over 2})} \right)
\left({\lambda_\alpha + i(\xi_- -{1\over 2}) \over
       \lambda_\alpha - i(\xi_- -{1\over 2})} \right)
\left({\lambda_\alpha + {i\over 2} \over
       \lambda_\alpha - {i\over 2}} \right)^{2N}
= & \prod_{\beta \ne \alpha}^M
\left( {\lambda_\alpha - \lambda_\beta + i \over
        \lambda_\alpha - \lambda_\beta - i} \right)
\left( {\lambda_\alpha + \lambda_\beta + i \over
        \lambda_\alpha + \lambda_\beta - i} \right) \,, \cr
& \qquad \alpha = 1, \cdots , M  \,.
\eqalignlabel{BA} \cr} $$
The corresponding energy eigenvalues are given by
$$E = - {1\over 2}\sum_{\alpha=1}^M {1 \over \lambda^2_\alpha + {1\over 4}}
 \eqlabel{energy} $$
(plus terms that are independent of $\{ \lambda_\alpha \}$), and
the corresponding $S^z$ eigenvalues are given by
$$S^z = {N\over 2} - M \,. \eqlabel{spin}  $$

The BA equations can be written more compactly as
$$ e_{2\xi_+ -1} (\lambda_\alpha)\ e_{2\xi_- -1} (\lambda_\alpha)\
e_1(\lambda_\alpha)^{2N+1} = - \prod_{\beta=1}^M
e_2(\lambda_\alpha - \lambda_\beta)\ e_2(\lambda_\alpha + \lambda_\beta)
\,, \eqlabel{BAcompact} $$
where
$$ e_n (\lambda) = {\lambda + {in\over 2}\over \lambda - {in\over 2}}
\,. \eqnum $$

We restrict the solutions of the BA equations as follows:
$${\hbox{ Re}} \left( \lambda_\alpha \right) \ge 0 \,,
\qquad \lambda_\alpha \ne 0 \,, \infty \,.  \eqlabel{restrict} $$
Within the coordinate Bethe Ansatz approach, these restrictions can be
understood by examining the Bethe wavefunction.
Alcaraz {\it et al.}${}^{\refref{alcaraz}}$ write the wavefunction in
terms of momenta $\{ k_\alpha \}$, which are related to our rapidities
$\{ \lambda_\alpha \}$ by
$\exp (i k_\alpha) = (\lambda_\alpha + {i\over 2})/
(\lambda_\alpha - {i\over 2})$. Because of the
periodicity $Re(k_\alpha) \rightarrow Re(k_\alpha) + 2\pi$ of the
wavefunction, one can make the restriction $-\pi \le Re(k_\alpha) \le \pi$.
Moreover, changing the sign of (any one) $k_\alpha$ results in a change of
sign of the wavefunction, and so does {\it not} lead to a new independent
Bethe state. Thus, we can make the further restriction
$0 \le Re(k_\alpha) \le \pi$. Finally, one can verify that the
wavefunction vanishes identically for $k_\alpha = 0 \,, \pi$.
Translating the restrictions $0 \le Re(k_\alpha) \le \pi \,, \quad
k_\alpha \ne 0 \,, \pi$ in terms of $\lambda_\alpha$ leads to
\eq\eqref{restrict} above. See also Refs. \refref{destri/devega},
\refref{alcaraz}, \refref{fendley/saleur}, \refref{hamer/quispel/batchelor},
\refref{hamer/batchelor}.

We adopt the ``string hypothesis'', which states that in the thermodynamic
($N \rightarrow \infty$) limit, all the solutions are collections of $M_n$
strings of length $n$ of the form (for $M_n > 0$)
$$\lambda_\alpha^{(n,j)} = \lambda_\alpha^n + i \left({n+1\over 2} - j \right)
\,, \eqlabel{string} $$
where $j = 1, \cdots , n$; $\alpha = 1, \cdots, M_n$;
$n = 1, \cdots, \infty$; and the centers $\lambda_\alpha^n$ are real and
non-negative. The total number of $\lambda$ variables is
$M = \sum_{n=1}^\infty n M_n$.

Implementing this hypothesis in the BA equations \eqref{BAcompact},
and then (following Takahashi${}^{\refref{takahashi1}}$ and
Gaudin${}^{\refref{gaudin}}$, and using the notation of Tsvelick and
Wiegmann${}^{\refref{tsvelick/wiegmann}}$) forming the product
$\prod_{j=1}^n$ over the imaginary parts of the strings, we obtain a set
of equations for the centers $\lambda_\alpha^n$:
$$\eqalignno{
& e_n(\lambda_\alpha^n)^{2N+1}
\prod_{l=1}^{d(n, 2\xi_+ -1)}e_{n+2\xi_+ - 2l} (\lambda^n_\alpha)
\prod_{l=1}^{d(n, 2\xi_- -1)}e_{n+2\xi_- - 2l} (\lambda^n_\alpha) \cr
& \qquad = (-)^n \prod_{m=1}^\infty \prod_{\beta=1}^{M_m}
E_{nm}(\lambda_\alpha^n - \lambda_\beta^m) \
E_{nm}(\lambda_\alpha^n + \lambda_\beta^m) \,,
\eqalignlabel{BAcenters} \cr} $$
where
$$E_{nm}(\lambda) = e_{|n-m|}(\lambda)\ e_{|n-m|+2}^2(\lambda)\ \cdots
\ e_{n+m-2}^2(\lambda)\ e_{n+m}(\lambda) \,,
\eqnum $$
and
$$d(n,x)=\left\{ \matrix{ min(n,x) &{\hbox{ if }} x & = {\hbox{ integer }}\cr
                            n &{\hbox{ if }} x & \ne {\hbox{ integer }} \cr}
\right. \eqnum $$
In obtaining the above result, we use the lemmas
$$ \prod_{j=1}^n e_x (\lambda_\alpha^{(n,j)}) = \prod_{l=1}^{d(n,x)}
e_{n+x+1-2l} (\lambda_\alpha^n) \,, \eqlabel{lemma1} $$
$$ \prod_{j=1}^n \prod_{k=1}^m  e_2 (\lambda_\alpha^{(n,j)}
- \lambda_\beta^{(m,k)}) = E_{nm} (\lambda_\alpha^n -\lambda_\beta^m) \,.
\eqlabel{lemma2} $$

Since the equations \eqref{BAcenters} involve only products of phases,
it is useful to take the logarithm:
$$\eqalignno{
& (2N+1) q_n(\lambda_\alpha^n)
+ \sum_{l=1}^{d(n, 2\xi_+ - 1)} q_{n+2\xi_+ - 2l} (\lambda^n_\alpha)
+ \sum_{l=1}^{d(n, 2\xi_- - 1)} q_{n+2\xi_- - 2l} (\lambda^n_\alpha) \cr
& \qquad = 2\pi J_\alpha^n
+ \sum_{m=1}^\infty \sum_{\beta=1}^{M_{m}}\left[
\Xi_{n m} (\lambda_\alpha^n - \lambda_\beta^{m})
+ \Xi_{n m} (\lambda_\alpha^n + \lambda_\beta^{m}) \right]
\,, \eqalignlabel{BAlog} \cr}$$
where $q_n (\lambda)$ are odd monotonic-increasing functions defined by
$$q_n (\lambda) = \pi +  i \ln e_n (\lambda) \,,
\qquad -\pi < q_n (\lambda) \le \pi \,, \eqnum $$
and $\Xi_{n m} (\lambda)$ are given by
$$ \Xi_{n m} (\lambda) = (1 - \delta_{n m}) q_{|n-m|}(\lambda)
+ 2q_{|n-m|+2}(\lambda)
+ \cdots + 2q_{n+m-2}(\lambda) + q_{n+m}(\lambda) \,. \eqnum $$
Moreover, $J_\alpha^n$ are integers in the range
$$J_{min}^n \le J_\alpha^n \le J_{max}^n \,, \eqlabel{range} $$
where, for the case $\xi_\pm = \infty$,
$$J_{min}^n = 1 \,, \qquad
  J_{max}^n = N + M_n - 2\sum_{m=1}^\infty min (m,n)\ M_m \,. \eqnum $$
There are similar formulas for the case $\xi_\pm \ne \infty$. (See also
Refs. \refref{alcaraz},\refref{hamer/batchelor}.)
We assume that the integers $\{ J_\alpha^n \}$ can be regarded as quantum
numbers of the model: for every set $\{ J_\alpha^n \}$ in the range
\eqref{range} (no two of which are identical),
there is a unique solution $\{ \lambda_\alpha^n \}$ (no two of which are
identical) of \eq\eqref{BAlog}.

For definiteness, we henceforth consider the case that
$2\xi_\pm$ is not an integer, and therefore, $d(n, 2\xi_\pm - 1) = n$
in \eq\eqref{BAlog}. Moreover, for later convenience, we
further restrict $\xi_\pm > 1/2$. (See \eq\eqref{definitions} below.)

Evidently, in order to calculate the boundary $S$ matrices, we
must investigate the low-lying excitations of the spin chain. However, as a
useful warm-up exercise, we first consider the ground state.
\vfill\eject

\noindent
{\bf \chapnum . Ground state}
\vskip 0.2truein

As discussed in Ref. \refref{alcaraz}, the ground state is an $S^z = 0$ state
characterized by a filled ``Fermi sea'' of strings of length 1.
That is, $M_1 = {N\over 2}$ and $M_n = 0$ for $n > 1$.
(We consider the case that $N$ is even.) We define the ``counting'' function
$h(\lambda)$ by
$$h(\lambda) = {1\over 2\pi} \left\{ (2N +1)q_1(\lambda)
+ q_{2\xi_+ -1}(\lambda) + q_{2\xi_- -1}(\lambda)
- \sum_{\beta = 1}^{N\over 2} \left[ q_2(\lambda - \lambda_\beta^1)
+ q_2(\lambda + \lambda_\beta^1) \right] \right\}
\,, \eqlabel{counting} $$
so that the BA equations \eqref{BAlog} for $n=1$ are
$$ h (\lambda_\alpha^1) = J_\alpha^1
\,, \qquad \alpha = 1\,, \cdots \,, {N\over 2} \,. \eqnum $$
We define the root density $\rho(\lambda)$ by
$$\rho(\lambda) = {1\over N} {d h (\lambda)\over d \lambda} \,.
\eqnum $$
In order to control terms of order $1/N$, one must exercise care
when passing from the sum in \eqref{counting} to an integral.
In particular, as shown in the Appendix (see \eq\eqref{appendix}),
there is an additional term of order $1/N$. Hence, the root density
obeys the integral equation
$$\eqalignno{
\rho(\lambda) &= 2 a_1(\lambda)
- \int_0^\infty d\lambda' \left[ a_2(\lambda - \lambda')
+ a_2(\lambda + \lambda') \right] \rho (\lambda') \cr
&\quad + {1\over N} \left[ a_1(\lambda)
+ a_2(\lambda) + a_{2\xi_+ -1}(\lambda) + a_{2\xi_- -1}(\lambda) \right]
\,, \qquad \lambda > 0  \eqalignlabel{intgeq} \cr}$$
(plus terms that are higher order in $1/N$), where
$$a_n(\lambda) = {1\over 2\pi} {d q_n (\lambda)\over d\lambda}
= {1\over 2\pi} {n\over \lambda^2 + {n^2\over 4}}
= {1\over 2\pi} \int_{-\infty}^\infty d\omega \
e^{-i \omega \lambda}\ e^{-n |\omega|/2} \,, \qquad n > 0 \,. \eqnum $$
In order to solve this equation, it is convenient
to extend the domain of the root density to negative values of $\lambda$,
such that it is an even function of $\lambda$. That is, we introduce
the symmetric density $\rho_s(\lambda)$
$$\rho_s(\lambda) = \left\{
\matrix{ \rho(\lambda) \quad \quad \lambda > 0 \cr
         \rho(-\lambda) \quad \quad \lambda < 0 \cr} \right.
\,. \eqlabel{symmetric}  $$
This density satisfies the linear integral equation
$$\rho_s = 2 a_1  - a_2 * \rho_s + {1\over N} \left( a_1 + a_2 + a_{2\xi_+ -1}
+ a_{2\xi_- -1} \right) \,, \eqnum $$
where $*$ denotes the convolution
$$ \left( f * g \right) (\lambda) = \int_{-\infty}^\infty
d\lambda'\ f(\lambda - \lambda') g(\lambda') \,. \eqnum $$
The solution is readily found by Fourier transforms, and is given by
$$\rho_s (\lambda) = 2 s(\lambda) + {1\over N} \left[ s(\lambda) +
J(\lambda) + J_+(\lambda) + J_-(\lambda) \right] \,,
\eqlabel{vacuumdensity} $$
where
$$\eqalignno{
s(\lambda) &= {1\over 2 \ch \pi \lambda}
= {1\over 2\pi} \int_{-\infty}^\infty d\omega\ e^{-i \omega \lambda}\
{e^{- |\omega|/2}\over 1 + e^{-|\omega|}} \,, \cr
J(\lambda) &= (a_1 * s) (\lambda)
= {1\over 2\pi} \int_{-\infty}^\infty d\omega\ e^{-i \omega \lambda}\
{e^{- |\omega|}\over 1 + e^{-|\omega|}} \,, \cr
J_\pm (\lambda) &= (a_{2\xi_\pm -2} * s) (\lambda)
= {1\over 2\pi} \int_{-\infty}^\infty d\omega\ e^{-i \omega \lambda}\
{e^{-(\xi_\pm - {1\over 2})|\omega|}\over 1 + e^{-|\omega|}}
\,. \eqalignlabel{definitions} \cr} $$

We calculate the ground state energy density $e_N^{vac} = E/N$ with the
help of \eq\eqref{energy}.
Again exercising care in passing from a sum to an integral, we obtain
(to order $1/N$)
$$\eqalignno{
e_N^{vac} &= - {\pi\over N} \sum_{\alpha=1}^{N\over 2} a_1 (\lambda_\alpha^1)
\cr
          &= - {\pi\over 2} \left[\int_{-\infty}^\infty d\lambda \
a_1(\lambda) \rho_s (\lambda) - {1\over N} a_1(0) \right] \cr
&= e_\infty^{vac} + {f\over N} \,,
\eqalignlabel{vacuumenergy} \cr} $$
where $e_\infty^{vac}$ is the energy density of the infinite chain,
and the surface energy $f$ of the chain is given by
$$ f = {\pi\over 2} \left\{ s(0)
- \int_{-\infty}^\infty d\lambda \
a_1(\lambda) \left[ s(\lambda) + J_+(\lambda) + J_-(\lambda) \right] \right\}
\,. \eqlabel{surfaceenergy} $$
Our results for the root density
\eqref{vacuumdensity} and the surface energy \eqref{surfaceenergy}
agree with those of Ref. \refref{hamer/quispel/batchelor}.

\vskip 0.4truein
\noindent
{\bf \chapnum . Excited states and $S$ matrix}
\vskip 0.2truein

As shown by Faddeev and Takhtajan${}^{\refref{faddeev/takhtajan}}$,
the isotropic antiferromagnetic Heisenberg chain with {\it periodic}
boundary conditions has spin $1/2$ excitations (``kinks'' or ``spinons'').
Following their approach, we consider the case that $N$ is even, and
therefore the number of such spinons is even.
Hence, the lowest-lying excited states have two spinons, and
there are four such states: the triplet $(S=1)$ and singlet $(S=0)$ states.

For the {\it open} spin chain, the bulk terms in the Hamiltonian are
$SU(2)$-invariant; hence, the excitations are still spin $1/2$ spinons,
and there are four degenerate states with two spinons.
However, the boundary terms break the $SU(2)$ symmetry, and so the
total spin is no longer a good quantum number. We now proceed to investigate
the excited states, which we classify by their $S^z$ eigenvalue.
(Excited states of the open spin $1/2$ chain have also been studied
in Refs. \refref{destri/devega}, \refref{hamer/batchelor}.)

\vskip 0.2truein
\noindent
$S^z = 1$ {\it state}
\vskip 0.2truein

The $S^z =1$ state is characterized by two holes in the sea of strings of
length 1, and no strings of greater length. That is,
$M_1 = {N\over 2}-1$ and $M_n = 0$ for $n > 1$. The counting function
$h^{(1)}(\lambda)$ is given by
$$h^{(1)}(\lambda) = {1\over 2\pi} \left\{ (2N +1)q_1(\lambda)
+ q_{2\xi_+ -1}(\lambda) + q_{2\xi_- -1}(\lambda)
- \sum_{\beta = 1}^{{N\over 2}-1} \left[ q_2(\lambda - \lambda_\beta^1)
+ q_2(\lambda + \lambda_\beta^1) \right] \right\}
\,. \eqnum $$
The hole rapidities $\tilde\lambda_1 \,, \tilde\lambda_2$ are given by
$$h^{(1)}(\tilde\lambda_\alpha) = \tilde J_\alpha \,, \qquad
\alpha = 1 \,, 2 \,, \eqlabel{first} $$
where $\tilde J_1 \,, \tilde J_2$ are integers in the range
$$J_{min}^1 \le \tilde J_1 \,, \tilde J_2 \le J_{max}^1 \,, \eqnum $$
and which are {\it not} in the set of integers $\{ J_\alpha^1 \}$
corresponding to $\{ \lambda_\alpha^1 \}$.

We define the quantity $\sigma^{(1)}(\lambda)$
$$\sigma^{(1)}(\lambda) = {1\over N} {d h^{(1)}(\lambda)\over d \lambda} \,,
\eqlabel{sigma1} $$
which is the density of roots {\it plus} the density of holes.
\footnote*{Since $\sigma^{(1)}$ depends also on the hole rapidities, the
notation $\sigma^{(1)}(\lambda \,, \tilde\lambda_1 \,, \tilde\lambda_2)$
would be more accurate. However, following the usual practice, we
suppress the dependence on the hole rapidities.}
We proceed as in the case of the ground state, being careful
when passing from the sum to an integral (see \eq\eqref{appendix}), and
we obtain the integral equation
$$\eqalignno{
\sigma^{(1)}(\lambda) &= 2 a_1(\lambda)
- \int_0^\infty d\lambda' \left[ a_2(\lambda - \lambda')
+ a_2(\lambda + \lambda') \right] \sigma^{(1)} (\lambda') \cr
& + {1\over N} \Big\{ a_1(\lambda)
+ a_2(\lambda) + a_{2\xi_+ -1}(\lambda) + a_{2\xi_- -1}(\lambda) \cr
& + \sum_{\alpha=1}^2 \left[a_2(\lambda - \tilde\lambda_\alpha)
+ a_2(\lambda + \tilde\lambda_\alpha) \right]
\Big\} \,, \qquad \lambda > 0 \eqalignnum \cr} $$
(plus terms that are higher order in $1/N$).
The symmetric density $\sigma_s^{(1)}(\lambda)$ defined by
$$\sigma_s^{(1)}(\lambda) = \left\{
\matrix{ \sigma^{(1)}(\lambda) \quad \quad \lambda > 0 \cr
         \sigma^{(1)}(-\lambda) \quad \quad \lambda < 0 \cr} \right.
\,, \eqlabel{sigma}  $$
is therefore given by
$$\sigma_s^{(1)} (\lambda) = 2 s(\lambda) + {1\over N} r^{(1)}(\lambda)
\,, \eqlabel{plus1density} $$
where $r^{(1)}(\lambda)$ is given by
$$r^{(1)}(\lambda) =  s(\lambda) + J(\lambda) + J_+(\lambda) + J_-(\lambda)
+ \sum_{\alpha=1}^2 \left[ J(\lambda - \tilde\lambda_\alpha)
+ J(\lambda + \tilde\lambda_\alpha) \right]  \eqlabel{r} $$
(plus terms that are higher order in $1/N$).
The functions $s(\lambda)\,, J(\lambda) \,,$ and $J_\pm(\lambda)$
are defined in \eq\eqref{definitions}.

The energy density $e_N = E/N$ is given by
$$e_N = e_N^{vac} + {1\over N} \sum_{\alpha=1}^2
\varepsilon(\tilde\lambda_\alpha) \,, \eqlabel{anotherenergy} $$
where $e_N^{vac}$ is the ground state energy density \eqref{vacuumenergy},
and the spinon energy $\varepsilon(\lambda)$ is given by
$$\varepsilon(\lambda)= \pi s(\lambda) \eqlabel{energyhole} $$
(plus terms that are higher order in $1/N$).

The spinons of the antiferromagnetic Heisenberg chain with {\it periodic}
boundary conditions have the same energy, and have momentum $p(\lambda)$
given by${}^{\refref{faddeev/takhtajan}}$
$$p(\lambda) = \tan^{-1} \sh (\pi\lambda) - {\pi\over 2} \,.
\eqnum $$
For the {\it open} spin chain, we {\it define} $p(\lambda)$ by this equation.
Note that
$${d p \over d\lambda} = 2 \pi s(\lambda)  \,. \eqlabel{momentum} $$

We come now to the calculation of the $S$ matrix.
As already noted in the Introduction, our approach is a generalization
of the Korepin-Andrei-Destri method
${}^{\refref{korepin}, \refref{andrei/destri}}$. From Eqs.
\eqref{sigma1}, \eqref{plus1density}, and \eqref{momentum}, we obtain
the relation
$${1\over N} {d h^{(1)}\over d \lambda} = {1\over \pi} {d p \over d\lambda}
+ {1\over N} r^{(1)}(\lambda) \,. \eqnum $$
Integrating with respect to $\lambda$, we obtain
$${2\pi\over N} h^{(1)}(\lambda)  = 2 p(\lambda)
+ {\pi\over N} \int_{-\infty}^\infty d\lambda'\
\epsilon(\lambda - \lambda')\ r^{(1)}(\lambda') + \pi  \,,
\eqnum $$
where $\epsilon(x)= {\hbox{ sign }} x = x/|x|$.
Finally, we evaluate this expression at the spinon
rapidity $\tilde\lambda_1$, recalling from \eq\eqref{first}
that $h(\tilde\lambda_1) = \tilde J_1$. Thus,
$${2\pi\over N} \tilde J_1  = 2 p(\tilde\lambda_1)
+ {\pi\over N} \int_{-\infty}^\infty d\lambda\
\epsilon(\tilde\lambda_1 - \lambda)\ r^{(1)}(\lambda) + \pi \,.
\eqlabel{anotherresult} $$

On the other hand, let us recall the quantization condition \eqref{q3}
for two particles with factorized scattering
on an interval of length $L$. For the spin chain,
the number of spins $N$ replaces $L$, and the ``particles'' are in
fact spin $1/2$ spinons of rapidities $\tilde\lambda_1$ and $\tilde\lambda_2$.
Thus, we have
$$e^{i 2 p(\tilde\lambda_1) N}\ R_{12}(\tilde\lambda_1 - \tilde\lambda_2)\
K_1(\tilde\lambda_1 \,, \xi_-)\
R_{21}(\tilde\lambda_1 + \tilde\lambda_2)\ K_1(\tilde\lambda_1 \,, \xi_+)
= 1 \,. \eqlabel{qq3} $$

The bulk 2-particle $S$ matrix is $SU(2)$-invariant, and is
given by${}^{\refref{faddeev/takhtajan}}$
$$R(\lambda) =
 {\Gamma \left({-i\lambda\over 2} + {1\over 2}\right)\over
  \Gamma \left({i\lambda\over 2} + {1\over 2}\right)}
{\Gamma \left({i\lambda\over 2} + 1\right) \over
\Gamma \left({-i\lambda\over 2} + 1\right)}
{\left( \lambda {\cal I} - i {\cal P} \right)\over ( \lambda - i )} \,,
\eqlabel{bulk1} $$
where ${\cal I}$ and ${\cal P}$ are the $4 \times 4$ identity and permutation
matrices, respectively. This matrix has the following form
$$R(\lambda) =  \left( \matrix{ a(\lambda) &0  &0  &0 \cr
                               0  & b(\lambda) & c(\lambda) & 0 \cr
                               0  & c(\lambda) & b(\lambda) & 0 \cr
                               0  & 0     & 0  & a(\lambda) \cr} \right)
\,, \qquad \eqlabel{bulk2}
$$
with $a(\lambda) = b(\lambda) + c(\lambda)$.

The $U(1)$ symmetry of the Hamiltonian's boundary terms implies that the
boundary $S$ matrix is of the form
$$K(\lambda \,, \xi) = \left( \matrix{ \alpha(\lambda \,, \xi) &0  \cr
                                     0  & \beta(\lambda \,, \xi) \cr} \right)
\,. \eqlabel{form} $$
Our task is to explicitly determine the matrix elements
$\alpha(\lambda \,, \xi)$ and $\beta(\lambda \,, \xi)$.

For the $S^z = 1$ state, the quantization condition \eqref{qq3}
implies
$$e^{i 2 p(\tilde\lambda_1) N} a(\tilde\lambda_1 - \tilde\lambda_2)\
\alpha(\tilde\lambda_1 \,, \xi_-)\ a(\tilde\lambda_1 + \tilde\lambda_2)\
\alpha(\tilde\lambda_1 \,, \xi_+) = 1 \,. \eqlabel{qq4} $$
In terms of the bulk $(\phi^{(1)})$ and boundary $(\varphi)$ phase shifts
defined by
$$a(\lambda) = e^{i \phi^{(1)}(\lambda)} \,, \qquad\qquad
  \alpha(\lambda \,, \xi) = e^{i \varphi(\lambda\,, \xi)} \,,
\eqlabel{phases} $$
respectively, the quantization condition \eqref{qq4} becomes
$${2 \pi\over N} m = 2 p(\tilde\lambda_1) + {1\over N} \Phi^{(1)} \,,
\eqlabel{q4} $$
where $m$ is an integer, and the total phase shift $\Phi^{(1)}$ is given by
$$\Phi^{(1)} = \phi^{(1)}(\tilde\lambda_1 - \tilde\lambda_2) +
         \phi^{(1)}(\tilde\lambda_1 + \tilde\lambda_2) +
\varphi(\tilde\lambda_1 \,, \xi_-) +
\varphi(\tilde\lambda_1 \,, \xi_+) \,. \eqlabel{total} $$

Comparing our Bethe Ansatz
result \eqref{anotherresult} with the quantization condition
\eqref{q4}, we see that the total phase shift $\Phi^{(1)}$ is given by
$$\Phi^{(1)} = \pi \int_{-\infty}^\infty d\lambda\
\epsilon(\tilde\lambda_1 - \lambda)\ r^{(1)}(\lambda) \,.
\eqlabel{phi} $$
It is convenient to evaluate the derivative of this expression,
$$\eqalignno{
{d \Phi^{(1)}\over d \tilde\lambda_1} &= 2\pi \Big[
s(\tilde\lambda_1) + J(\tilde\lambda_1) + 2 J(2\tilde\lambda_1)
+ J_+(\tilde\lambda_1) + J_-(\tilde\lambda_1) \cr
& \quad + J(\tilde\lambda_1 - \tilde\lambda_2)
+ J(\tilde\lambda_1 + \tilde\lambda_2) \Big] \,.  \eqalignnum \cr} $$
In obtaining this result, we have used \eq\eqref{r} for $r^{(1)}(\lambda)$;
and we have remembered to differentiate also $r^{(1)}(\lambda)$ in
\eq\eqref{phi}. The bulk phase shift $\phi^{(1)}(\lambda)$ given by
Eqs. \eqref{bulk1},\eqref{bulk2},\eqref{phases} satisfies
$${d \phi^{(1)}(\lambda)\over d\lambda} = 2\pi J(\lambda) \,. \eqnum $$
In view also of \eq\eqref{total}, we see that the boundary phase shift
$\varphi(\lambda \,, \xi_\pm)$ satisfies
$${d \varphi(\lambda \,, \xi_\pm)\over d\lambda} = \pi \left[
s(\lambda) + J(\lambda) + 2 J(2 \lambda) + 2 J_\pm(\lambda) \right]
\,. \eqlabel{main} $$
The phase shift is now readily evaluated by expressing the RHS
in terms of the Fourier transform, and using the identity
$$\int_0^\infty {d \omega\over \omega}
{e^{-\nu \omega}\over 1 + e^{-\omega}} = \ln \left(
{\Gamma \left({\nu\over 2}\right)\over \Gamma \left({\nu +1\over 2}\right)}
\right) \,, \qquad {\hbox{  Re }}\nu > 0 \,, \eqnum $$
and also the duplication formula for the gamma function,
$$2^{2z - 1} \Gamma(z)\ \Gamma(z + {1\over 2}) = \pi^{1\over 2} \Gamma(2z)
\,. \eqnum $$
The result for the boundary $S$ matrix element (up to a multiplicative
constant) is
$$\alpha(\lambda \,, \xi_\pm) = e^{i \varphi(\lambda\,, \xi_\pm)} =
{\Gamma \left({-i\lambda\over 2} + {1\over 4}\right) \over
  \Gamma \left({i\lambda\over 2} + {1\over 4}\right)}
{\Gamma \left({i\lambda\over 2} + 1\right) \over
\Gamma \left({-i\lambda\over 2} + 1\right)}
{\Gamma \left({-i\lambda\over 2} + {1\over 4}(2\xi_\pm -1)\right)\over
  \Gamma \left({i\lambda\over 2} + {1\over 4}(2\xi_\pm -1)\right)}
{\Gamma \left({i\lambda\over 2} + {1\over 4}(2\xi_\pm +1)\right)\over
\Gamma \left({-i\lambda\over 2} + {1\over 4}(2\xi_\pm +1)\right)} \,.
\eqlabel{result1} $$

\vskip 0.2truein
\noindent
$S^z = -1$ {\it state}
\vskip 0.2truein

To determine the remaining element
$\beta(\lambda \,, \xi)$ of the boundary $S$ matrix, we consider the
$S^z = -1$ state. This state is most easily described by changing
the pseudovacuum. Hence, instead of working with the states
\eqref{state}, we work now with
$${\cal C}(\lambda_1)\ {\cal C}(\lambda_2) \cdots {\cal C}(\lambda_M)\
\omega^- \,, \eqnum $$
where $\omega^-$ is the ferromagnetic vacuum state with all spins down,
$$ {\cal B}(\lambda)\ \omega^- = 0 \,. \eqlabel{pseudovacuum} $$
Sklyanin has shown${}^{\refref{sklyanin}}$ that
$\{ \lambda_\alpha \}$ satisfy the same BA equations \eqref{BA}
as before, except for the replacement of $\xi_\pm$ by $-\xi_\pm$.
The energy eigenvalues are given by the same expression
\eqref{energy}, and the $S^z$ eigenvalues are now given by
$$S^z = M - {N\over 2} \,. \eqnum  $$

The $S^z = -1$ state now consists of two holes in the sea of strings of
length 1, and no strings of greater length; i.e.,
$M_1 = {N\over 2}-1$ and $M_n = 0$ for $n > 1$. The calculation of the
density of roots plus holes is exactly the same as for the $S^z = 1$
state, except that we must track the change
$\xi_\pm \rightarrow -\xi_\pm$. We find that the density is given by
Eqs. \eqref{plus1density} and \eqref{r}, except that
$J_\pm(\lambda)$ is now given by
$$J_\pm (\lambda) = - (a_{2\xi_\pm} * s) (\lambda)
= -{1\over 2\pi} \int_{-\infty}^\infty d\omega\ e^{-i \omega \lambda}\
{e^{-(\xi_\pm + {1\over 2})|\omega|}\over 1 + e^{-|\omega|}}\,.
\eqlabel{jnew} $$
Moreover, the quantization condition
\eqref{qq3} and the form \eqref{bulk2},\eqref{form} of the $S$ matrices
imply
$$e^{i 2 p(\tilde\lambda_1) N} a(\tilde\lambda_1 - \tilde\lambda_2)\
\beta(\tilde\lambda_1 \,, \xi_-)\ a(\tilde\lambda_1 + \tilde\lambda_2)\
\beta(\tilde\lambda_1 \,, \xi_+) = 1 \,. \eqnum $$
Introducing the phase shift $\psi(\lambda\,, \xi)$ by
$\beta(\lambda \,, \xi) = \exp (i \psi(\lambda\,, \xi))$, we find that
$\psi(\lambda\,, \xi_\pm)$ obeys \eq\eqref{main}, with $J_\pm$ given by
\eq\eqref{jnew}. We conclude that the element $\beta(\lambda\,, \xi)$
of the boundary $S$ matrix is given (up to a multiplicative
constant) by
$$\beta(\lambda\,, \xi_\pm) = -{\lambda +i(\xi_\pm -{1\over 2})\over
\lambda -i(\xi_\pm -{1\over 2})} \alpha(\lambda) \,, \eqlabel{result2} $$
where $\alpha(\lambda)$ is given by \eq\eqref{result1}.
This completes the derivation of the result \eqref{resulta}, \eqref{resultb}
for the boundary $S$ matrix.

\vskip 0.2truein
\noindent
$S^z = 0$ {\it states}
\vskip 0.2truein

We have already succeeded to determine the boundary $S$ matrix. Nevertheless,
a good check on this result and on the general formalism is provided by
analyzing the $S^z = 0$ states. In particular, with the help of
expressions for the densities, we can determine the ``momentum''
$p(\tilde\lambda_1)$ of a spinon in terms of its deviation from the
free-particle value. Substituting this result, as well as the known
results for $R(\lambda)$ and $K(\lambda \,, \xi_\pm)$,
into the quantization condition \eqref{qq3}, we obtain a consistency condition.

To this end, we consider the $S^z = 0$ state consisting of two holes in the
sea of strings of length 1, and also one string of length 2; i.e.,
$M_1 = {N\over 2} - 2$, $M_2 = 1$,  and $M_n = 0$ for $n > 2$. For
$\xi_\pm \rightarrow \infty$, this is the spin-singlet $(S = S^z = 0)$
state. The counting function $h^{(0)}(\lambda)$ is given by
(see the BA equation \eqref{BAlog} with $n=1$)
$$\eqalignno{
h^{(0)}(\lambda) &= {1\over 2\pi} \Big\{ (2N +1)q_1(\lambda)
+ q_{2\xi_+ -1}(\lambda) + q_{2\xi_- -1}(\lambda)
- \sum_{\beta = 1}^{{N\over 2} - 2} \left[ q_2(\lambda - \lambda_\beta^1)
+ q_2(\lambda + \lambda_\beta^1) \right] \cr
& - q_1(\lambda - \lambda_0) - q_3(\lambda - \lambda_0)
  - q_1(\lambda + \lambda_0) - q_3(\lambda + \lambda_0) \Big\}
\,, \eqalignnum \cr} $$
where $\lambda_0 \equiv \lambda^2_1$ is the position of the center of the
string of length 2. The density $\sigma^{(0)}(\lambda)$ (defined
analogously as in \eq\eqref{sigma1})
obeys the integral equation
$$\eqalignno{
\sigma^{(0)}(\lambda) &= 2 a_1(\lambda)
- \int_0^\infty d\lambda' \left[ a_2(\lambda - \lambda')
+ a_2(\lambda + \lambda') \right] \sigma^{(0)} (\lambda') \cr
& + {1\over N} \Big\{ a_1(\lambda)
+ a_2(\lambda) + a_{2\xi_+ -1}(\lambda) + a_{2\xi_- -1}(\lambda) \cr
& + \sum_{\alpha=1}^2 \left[a_2(\lambda - \tilde\lambda_\alpha)
+ a_2(\lambda + \tilde\lambda_\alpha) \right] \cr
& - a_1(\lambda - \lambda_0) - a_3(\lambda - \lambda_0)
  - a_1(\lambda + \lambda_0) - a_3(\lambda + \lambda_0)
\Big\} \,, \qquad \lambda > 0 \eqalignnum \cr} $$
(plus terms that are higher order in $1/N$). We conclude that
the symmetric density $\sigma_s^{(0)}(\lambda)$ (defined analogously as in
\eq\eqref{sigma}) is given by
$$\sigma_s^{(0)} (\lambda) = 2 s(\lambda) + {1\over N}r^{(0)}(\lambda) \,,
\eqlabel{ddensity} $$
where
$$r^{(0)}(\lambda) = r^{(1)}(\lambda)
- \left[ a_1(\lambda - \lambda_0)
+ a_1(\lambda + \lambda_0) \right] \,, \eqlabel{density} $$
and $r^{(1)}(\lambda)$ is given in \eq\eqref{r}.

The position $\lambda_0$ of the center of the string of length 2
can be determined in terms of the positions $\tilde\lambda_1$ and
$\tilde\lambda_2$ of the holes. We accomplish this using the BA
equations \eqref{BAlog} with $n=2$:
$$\eqalignno{
& (2N+1) q_2(\lambda_0)
+ q_{2\xi_+} (\lambda_0) + q_{2\xi_+ - 2} (\lambda_0)
+ q_{2\xi_-} (\lambda_0)  + q_{2\xi_- - 2} (\lambda_0) \cr
& \qquad = 2\pi J_1^2
+ \sum_{\beta=1}^{{N\over 2} - 2}\left[
\Xi_{2 1} (\lambda_0 - \lambda_\beta^1)
+ \Xi_{2 1} (\lambda_0 + \lambda_\beta^1) \right] +  2 q_1(\lambda_0)
+ q_2(\lambda_0) \,. \eqalignlabel{BAposition} \cr}$$
(We remind the reader that, in terms of the notation in \eq\eqref{BAlog},
$\lambda_0$ is in fact $\lambda_1^2$.)
Passing again from the sum to an integral, and making use of the
result \eqref{ddensity}, we are led to the constraint
$$ e_{2\xi_+ -2}(\lambda_0)\
   e_{2\xi_- -2}(\lambda_0)\
   e_1(\lambda_0 - \tilde\lambda_1)\
   e_1(\lambda_0 - \tilde\lambda_2)\
   e_1(\lambda_0 + \tilde\lambda_1)\
   e_1(\lambda_0 + \tilde\lambda_2) = 1 \,.
\eqlabel{constraint} $$
This is considerably more complicated than the corresponding
constraint for the {\it periodic} chain, which implies
$\lambda_0^{ \ (periodic)} = (\tilde\lambda_1 + \tilde\lambda_2)/2$.

For the case $\xi_\pm = \infty$, the constraint becomes
$$ e_1(\lambda_0 - \tilde\lambda_1)\
   e_1(\lambda_0 - \tilde\lambda_2)\
   e_1(\lambda_0 + \tilde\lambda_1)\
   e_1(\lambda_0 + \tilde\lambda_2) = 1 \,. \eqnum $$
In addition to the solution $\lambda_0 = 0$
\footnote*{The restriction \eqref{restrict} can presumably be strengthened
to Re $(\lambda_\alpha) > 0$, and thus, we discard this solution.},
this constraint has the solution
$$ \lambda_0 = {\sqrt{ {1\over 4} + {1\over 2}\left[
(\tilde\lambda_1)^2 + (\tilde\lambda_2)^2 \right] }} \,.
\eqlabel{center} $$
For $\xi_\pm \ne \infty$, \eq\eqref{constraint} for $\lambda_0$
is not difficult to solve. Nevertheless, the solution is given by a rather
cumbersome expression, which we shall not present here.

One can verify that this $S^z = 0$ state has the same energy
\eqref{anotherenergy}, \eqref{energyhole} as the $S^z = 1$ state.
We now express the spinon ``momentum'' $p(\tilde\lambda_1)$ in terms of its
deviation from the free-particle value. In analogy with the $S^z = 1$ case,
we have
$${2\pi\over N} \tilde J_1  = 2 p(\tilde\lambda_1)
+ {\pi\over N} \int_{-\infty}^\infty d\lambda\
\epsilon(\tilde\lambda_1 - \lambda)\ r^{(0)}(\lambda) + \pi \,. \eqnum $$
Substituting for $r^{(0)}(\lambda)$ the expression \eqref{density},
we obtain the result
$$2 p(\tilde\lambda_1) = {2\pi\over N} \tilde J_1
+ {1\over N}\left[ -\Phi^{(1)}
 +q_1 (\tilde\lambda_1 - \lambda_0) + q_1 (\tilde\lambda_1 + \lambda_0)
\right] \,, \eqlabel{qq} $$
where $\Phi^{(1)}$ is given by \eq\eqref{phi}.

We turn now to the quantization condition.
For the $S^z = 0$ states, the quantization condition \eqref{qq3}
leads to a $2 \times 2$ matrix equation. The two eigenvalues of this matrix
are pure phases. (Since the matrix elements of $R(\lambda)$ and
$K(\lambda \,, \xi_\pm)$ are known, these eigenvalues can be computed
explicitly. However, for $\xi_\pm \ne \infty$, the actual expressions
for the eigenvalues are cumbersome, and we do not give them here.)
We consider the eigenvalue $\exp i\Phi^{(0)}$ which
for $\xi_\pm \rightarrow \infty$ corresponds to the spin-singlet
$(S = S^z = 0)$ state.\footnote{$\dagger$}{We implicitly assume that
$\Phi^{(0)}$ is a continuous function of $\xi_\pm$.}
Hence, the quantization condition implies
$$e^{i 2 p(\tilde\lambda_1) N}  e^{i\Phi^{(0)}} = 1 \,.
\eqlabel{totalphasesinglet} $$

Substituting the result for $2 p(\tilde\lambda_1)$ given in \eq\eqref{qq}
into \eq\eqref{totalphasesinglet}, we obtain the consistency condition
$$ e^{i\left( \Phi^{(0)} - \Phi^{(1)} \right)}
= e_1 (\tilde\lambda_1 - \lambda_0)\
  e_1 (\tilde\lambda_1 + \lambda_0)
\,, \eqlabel{identity2}   $$
where $\tilde\lambda_1$ and $\tilde\lambda_2$ are arbitrary, and
$\lambda_0$ is a solution of \eq\eqref{constraint}.

This relation is readily verified for $\xi_\pm = \infty$. Indeed,
for this case, the boundary $S$ matrix is
proportional to the identity matrix $( \alpha(\lambda \,, \infty) =
\beta(\lambda \,, \infty) )$. The $2 \times 2$ matrix equation
which follows from the quantization condition is readily
diagonalized, and the phase shift $\Phi^{(0)}$ is found to be
$$
\Phi^{(0)} = \phi^{(0)}(\tilde\lambda_1 - \tilde\lambda_2)
+ \phi^{(0)}(\tilde\lambda_1 + \tilde\lambda_2) +
2 \varphi(\tilde\lambda_1 \,, \infty)  \,. \eqlabel{phiinf} $$
Here $\phi^{(0)}(\lambda)$ is the singlet phase shift
$$b(\lambda) - c(\lambda) = e^{i \phi^{(0)}(\lambda)}
= -\left( {\lambda + i\over \lambda - i} \right)
e^{i \phi^{(1)}(\lambda)} \,, \eqnum $$
where $\phi^{(1)}(\lambda)$ is the triplet phase shift.
(See Eqs. \eqref{bulk1},\eqref{bulk2},\eqref{phases}.) Since
$$\Phi^{(1)} = \phi^{(1)}(\tilde\lambda_1 - \tilde\lambda_2)
+ \phi^{(1)}(\tilde\lambda_1 + \tilde\lambda_2) +
2 \varphi(\tilde\lambda_1 \,, \infty)  \,, \eqnum $$
it follows that
$$ e^{i\left( \Phi^{(0)} - \Phi^{(1)} \right)} =
e_1 \left( {1\over 2}(\tilde\lambda_1 - \tilde\lambda_2) \right)\
e_1 \left( {1\over 2}(\tilde\lambda_1 + \tilde\lambda_2) \right)
\,. \eqlabel{deltaresult} $$
The relation \eqref{identity2} now follows from the algebraic identity
$$
e_1 (\tilde\lambda_1 - \lambda_0)\
e_1 (\tilde\lambda_1 + \lambda_0) =
e_1 \left( {1\over 2}(\tilde\lambda_1 - \tilde\lambda_2) \right)\
e_1 \left( {1\over 2}(\tilde\lambda_1 + \tilde\lambda_2) \right)
 \,. \eqlabel{identity}  $$
This identity is true for arbitrary values of $\tilde\lambda_1$
and $\tilde\lambda_2$, where $\lambda_0$ is given by \eqref{center}.

We have explicitly verified the formula \eqref{identity2}
also for the case $\xi_- = \infty$, $\xi_+ \ne \infty$, and presumably it is
true in general. This equality provides a nontrivial consistency
check of the bulk and boundary $S$ matrices and of the general
formalism.

\vskip 0.4truein
\noindent
{\bf \chapnum . Discussion}
\vskip 0.2truein

We have demonstrated how the boundary $S$ matrix for the open
Heisenberg chain with boundary magnetic fields can be calculated directly
from the Bethe Ansatz equations. As yet, this is the only quantum-mechanical
calculation of the boundary $S$ matrix.
We have restricted our attention in this paper to the case
that the bulk terms are $SU(2)$ invariant and the boundary terms are
$U(1)$ invariant only for the sake of simplicity. We see no difficulty
in extending our calculations to the case that both the bulk terms and
the boundary terms are only $U(1)$ invariant. However, the case where the
boundary terms have no continuous symmetry cannot be undertaken by this
approach until the corresponding Bethe Ansatz solution is
found${}^{\refref{ddevega/ruiz}}$.

The analysis of the $S^z=0$ states for the open chain differs
significantly from that of the periodic chain. Indeed, for the open
chain, we have seen that the position of the center of the string of
length 2 is a complicated function of the hole positions $\tilde\lambda_1$
and $\tilde\lambda_2$ (as well as the boundary parameters $\xi_\pm$);
while for the periodic chain, the center of the string is located
midway between the two holes. Naively, one might worry that this
leads to a breakdown of factorization. However, we have seen that
factorization is maintained by virtue of certain nontrivial identities.
We expect that a similar situation holds for the periodic chain
with four or more holes.

\bigskip

This work was supported in part by the National Science Foundation under
Grants PHY-92 22318 and PHY-92 09978.

\bigskip

\chapno = -1

\vskip 0.4truein
\noindent
{\bf \chapnum . Appendix}
\vskip 0.2truein

We derive in this Appendix a formula for approximating a sum of the form
$${1\over N}\sum_{\alpha=1}^{M_1} g(\lambda_\alpha^1) \eqlabel{sum} $$
by an integral, to order $1/N$, for large $N$.
Here $g(\lambda)$ is an arbitrary function of $\lambda$ which goes to
$0$ for $\lambda \rightarrow \infty$. Furthermore,
$\{ \lambda_\alpha^1 \}$ are solutions
of the BA equations \eqref{BAlog} with $n=1$. That is,
$$ h (\lambda_\alpha^1) = J_\alpha^1
\,, \qquad \alpha = 1\,, \cdots \,, M_1 \,, \eqnum $$
where $h(\lambda)$ is the appropriate counting function, and $M_1 \sim N/2$.
There are two key points in our discussion: (1) we use the
Euler-Maclaurin formula for approximating sums by integrals; and (2) we
use the fact that the solution $\lambda = 0$ of the BA equations is excluded.
The Euler-Maclaurin formula has been widely used to calculate finite-size
corrections in Bethe Ansatz systems. (See, e.g.,
Refs. \refref{hamer/quispel/batchelor}, \refref{finite}.)

Let us consider the case that there are $\nu$ holes, with
rapidities $\{ \tilde \lambda_\alpha \}$
$$ h (\tilde \lambda_\alpha) = \tilde J_\alpha
\,, \qquad \alpha = 1\,, \cdots \,, \nu \,. \eqnum $$
We denote the union of the sets $\{ \lambda_\alpha^1 \}$ and
$\{ \tilde \lambda_\alpha \}$ by $\{ \Lambda_\alpha \}$, with
$\alpha = 1\,, \cdots \,, I$, where $I = M_1 + \nu$.
Similarly, we denote the union of the sets $\{ J_\alpha^1 \}$ and
$\{ \tilde J_\alpha \}$ by $\{ j_\alpha \}$. Evidently,
$$ h (\Lambda_\alpha) = j_\alpha = \alpha
\,, \qquad \alpha = 1\,, \cdots \,, I \,. \eqnum $$
As noted in \eq\eqref{restrict}, $\Lambda_\alpha > 0$. Nevertheless,
it is useful to introduce $\Lambda_0 \equiv 0$. Since $h(0) = 0$, the
corresponding integer is $j_0 = 0$.

Instead of evaluating the sum \eqref{sum} directly, it is convenient
to first evaluate the sum
$${1\over N}\sum_{\alpha=0}^{I} g(\Lambda_\alpha) \,.
\eqlabel{sum2} $$
We approximate this sum using the Euler-Maclaurin formula (see, e.g.,
Ref. \refref{whittaker/watson})
$$\sum_{k=0}^n f(a + k \delta) = {1\over \delta} \int_a^{a + n \delta}
dx\ f(x) + {1\over 2}\left[ f(a) + f(a + n \delta) \right] + \cdots
\,. \eqnum $$
Indeed, transforming to a sum over equidistant points by means
of the change of variables $h(\lambda) = j$, we obtain
$$\eqalignno{
{1\over N}\sum_{\alpha=0}^{I} g(\Lambda_\alpha) &=
{1\over N}\sum_{\alpha=0}^{I} g \left( h^{\ -1} (j_\alpha) \right) \cr
&= {1\over N}\int_{j_0}^{j_{I}} dj \ g \left( h^{\ -1} (j) \right)
+ {1\over 2N} \left[ g \left( h^{\ -1} (j_0) \right)
+g \left( h^{\ -1} (j_{I}) \right) \right]
\eqalignnum \cr} $$
(plus terms that are higher order in $1/N$).
Introducing the density $\sigma(\lambda)$
$$\sigma(\lambda) = {1\over N} {d h (\lambda)\over d \lambda} \,,
\eqnum $$
we obtain
$${1\over N}\sum_{\alpha=0}^{I} g(\Lambda_\alpha) =
\int_{0}^{\Lambda_{I}} d\lambda \  \sigma(\lambda)\
g(\lambda) + {1\over 2N} \left[ g(0)
+ g(\Lambda_{I}) \right] \,. \eqnum $$
By definition of $\{ \Lambda_\alpha \}$,
$$\sum_{\alpha=0}^{I} g(\Lambda_\alpha) = g(0) +
\sum_{\alpha=1}^{M_1} g(\lambda_\alpha^1) +
\sum_{\alpha=1}^{\nu} g(\tilde\lambda_\alpha) \,. \eqnum
$$
Therefore,
$${1\over N}\sum_{\alpha=1}^{M_1} g(\lambda_\alpha^1)
= \int_0^{\Lambda_{I}} d\lambda \  \sigma(\lambda)\ g(\lambda)
- {1\over N}\sum_{\alpha=1}^{\nu} g(\tilde\lambda_\alpha)
- {1\over 2N} g(0) + {1\over 2N} g(\Lambda_{I})  \,.
\eqnum $$
Making the approximation $\Lambda_{I} \approx \infty$
introduces an error which is higher order in $1/N$. We conclude
that the sum \eqref{sum} is given by
$${1\over N}\sum_{\alpha=1}^{M_1} g(\lambda_\alpha^1)
= \int_0^\infty d\lambda \  \sigma(\lambda)\ g(\lambda)
- {1\over N}\sum_{\alpha=1}^{\nu} g(\tilde\lambda_\alpha)
- {1\over 2N} g(0)   \eqlabel{appendix} $$
(plus terms that are higher order in $1/N$). The presence of
the last term $-{1\over 2N} g(0)$ should be noted.

\vskip 0.4truein
\noindent
{\bf References}
\vskip 0.2truein

\reflabel{zamolodchikov/zamolodchikov}
A.B. Zamolodchikov and Al.B. Zamolodchikov, Ann. Phys. {\it 120} (1979) 253;
A.B. Zamolodchikov, Sov. Sci. Rev. {\it A2} (1980) 1.

\reflabel{reviews}
P.P. Kulish and E.K. Sklyanin, J. Sov. Math. {\it 19} (1982) 1596;
{\it Lecture Notes in Physics}, Vol. 151 (Springer, 1982) 61;
M. Jimbo, Int. J. Mod. Phys. {\it A4} (1989) 3759.

\reflabel{cherednik}
I.V. Cherednik, Theor. Math. Phys. {\it 61} (1984) 977.

\reflabel{sklyanin}
E.K. Sklyanin, J. Phys. {\it A21} (1988) 2375.

\reflabel{nonsymmetric}
L. Mezincescu and R.I. Nepomechie, J. Phys. {\it A24} (1991) L17.

\reflabel{spin1/fusion}
L. Mezincescu, R.I. Nepomechie and V. Rittenberg, Phys. Lett. {\it A147}
(1990) 70; L. Mezincescu and R.I. Nepomechie, J. Phys. {\it A25} (1992) 2533.

\reflabel{general}
L. Mezincescu and R.I. Nepomechie, Int. J. Mod. Phys. {\it A6} (1991) 5231;
Addendum, {\it A7} (1992) 5657.

\reflabel{analytical}
L. Mezincescu and R.I. Nepomechie, Nucl. Phys. {\it B372} (1992) 597.

\reflabel{destri/devega}
C. Destri and H.J. de Vega, Nucl. Phys. {\it B374} (1992) 692;
{\it B385} (1992) 361.

\reflabel{devega/ruiz}
H.J. de Vega and A. Gonz\'alez-Ruiz, J. Phys. {\it A26} (1993) L519;
preprints LPTHE-PAR 93-38; LPTHE-PAR 94-12

\reflabel{ddevega/ruiz}
H.J. de Vega and A. Gonz\'alez-Ruiz, preprint LPTHE-PAR 93-29

\reflabel{yue}
R.-H. Yue and Y.-X. Chen, J. Phys. {\it A26} (1993) 2989;
B.-Y. Hou and R.-H. Yue, Phys. Lett. {\it A183} (1993) 169;
R.-H. Yue, J. Phys. {\it A27} (1994) 1633.

\reflabel{karowski/zapletal}
M. Karowski and A. Zapletal, preprint (1993)

\reflabel{ruiz}
A. Gonz\'alez-Ruiz, preprint F.T./U.C.M.-94/1

\reflabel{kulish}
P.P. Kulish, in {\it Proceedings of the Wigner Symposium}, ed. V. Dobrev
and H-D. Doebner

\reflabel{fring/koberle}
A. Fring and R. K\"oberle, preprints USP-IFQSC/TH/93-06; USP-IFQSC/TH/93-12;
USP-IFQSC/TH/94-03

\reflabel{sasaki}
R. Sasaki, preprint YITP/U-93-33;
E. Corrigan, P.Dorey, R.H. Rietdijk, R. Sasaki, preprint YITP/U-94-11.

\reflabel{ghoshal/zamolodchikov}
S. Ghoshal and A. B. Zamolodchikov, preprint RU-93-20

\reflabel{ghoshal}
S. Ghoshal, preprints RU-93-51; RU-94-02

\reflabel{chim}
L. Chim, preprint RU-94-33

\reflabel{alcaraz}
F.C. Alcaraz, M.N. Barber, M.T. Batchelor, R.J. Baxter and G.R.W. Quispel,
J. Phys. {\it A20} (1987) 6397. See also M. Gaudin, Phys. Rev.
{\it A4} (1971) 386; {\it La fonction d'onde de Bethe} (Masson, 1983).

\reflabel{korepin}
V.E. Korepin, Theor. Math. Phys. {\it 76} (1980) 165;
V.E. Korepin, G. Izergin and N.M. Bogoliubov, {\it Quantum Inverse
Scattering Method, Correlation Functions and Algebraic Bethe Ansatz}
(Cambridge University Press, 1993).

\reflabel{andrei/destri}
N. Andrei and C. Destri, Nucl. Phys. {\it B231} (1984) 445.

\reflabel{fendley/saleur}
P. Fendley and H. Saleur, preprint USC-94-001.

\reflabel{faddeev/takhtajan}
L.D. Faddeev and L.A. Takhtajan, J. Sov. Math. {\it 24}, 241 (1984).

\reflabel{hamer/quispel/batchelor}
C.J. Hamer, G.R.W. Quispel and M.T. Batchelor, J. Phys. {\it A20} (1987) 5677.
See also A.L. Owczarek and R.J. Baxter, J. Phys. {\it A22} (1989) 1141;
M.T. Batchelor and C.J. Hamer, J. Phys. {\it A23} (1990) 761.

\reflabel{affleck}
I. Affleck, preprint UBCTP-93-25

\reflabel{hamer/batchelor}
C.J. Hamer and M.T. Batchelor, J. Phys. {\it A21} (1988) L173;

\reflabel{takahashi1}
M. Takahashi, Prog. Theor. Phys. {\it 46} (1971) 401.

\reflabel{gaudin}
M. Gaudin, Phys. Rev. Lett. {\it 26} (1971) 1301

\reflabel{tsvelick/wiegmann}
A.M. Tsvelick and P.B. Wiegmann, Adv. in Phys. {\it 32} (1983) 453.

\reflabel{finite}
R. Giles, L.D. McLerran, and C.B.Thorn, Phys. Rev. {\it D17} (1978) 2058;
F. Woynarovich and H.P. Eckle, J. Phys. {\it A20} (1987) L97;
M. Karowski, Nucl. Phys. {\it B300} (1988) 473.

\reflabel{whittaker/watson}
E.T. Whittaker and G.N. Watson, {\it A Course of Modern Analysis}
(Cambridge University Press, 1988).

\end